\documentclass[twocolumn,aps,prl,superscriptaddress,amssymb]{revtex4}
\usepackage{natbib}
\usepackage{graphicx}
\usepackage{units}
\usepackage{braket}
\usepackage{textcomp}
\usepackage{psfrag}
\usepackage{color}
\begin{document}

\hyphenation{Ryd-berg}

%\preprint{}

\title{Evidence for coherent collective Rydberg excitation in the strong blockade regime}


\author{Rolf Heidemann}
\email[Electronic address: ]{r.heidemann@physik.uni-stuttgart.de}
\author{Ulrich Raitzsch}
\author{Vera Bendkowsky}
\author{Bj\"{o}rn Butscher}
\author{Robert L\"{o}w}
\affiliation{5. Physikalisches Institut, Universit\"{a}t
Stuttgart, Pfaffenwaldring 57, 70569 Stuttgart, Germany}
\author{Luis Santos}
\affiliation{Institut f\"{u}r Theoretische Physik, Universit\"{a}t
Hannover, Appelstra{\ss}e 2, 30167 Hannover, Germany}
\author{Tilman Pfau}
\email[Electronic address:]{t.pfau@physik.uni-stuttgart.de}
\affiliation{5. Physikalisches Institut, Universit\"{a}t
Stuttgart, Pfaffenwaldring 57, 70569 Stuttgart, Germany}

%Rolf Heidemann, Ulrich Raitzsch, Vera Bendkowsky, Bj\"{o}rn Butscher, Robert L\"{o}w, Luis Santos and Tilman Pfau
%\email[Electronic address:
%\\]{r.heidemann@physik.uni-stuttgart.de}
%\thanks{}


\date{June 19, 2007}

\begin{abstract}
% abstract here

Blockade effects on the single quantum level are at the heart of
quantum devices like single-electron transistors. The blockade
mechanisms are based on strong interactions like the Coulomb
interaction in case of single electrons. Neutral atoms excited
into a Rydberg state experience abnormally strong interactions
that lead to the corresponding blockade effect for Rydberg atoms.
In this paper we report on our measurements of a strong van der
Waals blockade, showing that only one out of several thousand
atoms within a blockade volume can be excited. In addition, our
experimental results demonstrate the coherent nature of the
excitation of magnetically trapped ultracold atoms into a Rydberg
state, confirming the predicted dependence of the collective Rabi
frequency on the square root of the mesoscopic system size. This
collective coherent behaviour is generic for all mesoscopic
systems which are able to carry only one single excitation
quantum.
\end{abstract}

\maketitle

% body of paper here

Early studies on atomic beams, where line broadening effects at
high Rydberg densities were observed \citep{Raimond:1981},
triggered experiments on ultracold samples \citep{Tong:2004,
Singer:2004,Liebisch:2005,Anderson:1998,
Vogt:2006,Afrousheh:2006,Mourachko:1998,Carroll:2006,Cubel:2005,Deiglmayr:2006},
where the atomic motion of the atoms during the lifetime of the
Rydberg atoms can be neglected. This excited state of matter is
known as frozen Rydberg gas.  The coherent elastic interaction
between Rydberg atoms leads to a blockade effect which has been
proposed as a crucial ingredient for rapid quantum gates either
using single neutral atoms \cite{Jaksch:2000} or mesoscopic
samples \cite{Lukin:2001} to store and process quantum
information. This blockade effect has been studied in various
experiments using laser cooled atoms prepared in magneto-optical
traps \cite{Tong:2004, Singer:2004,Liebisch:2005}. Typically the
interaction effect was studied by changing the density of Rydberg
atoms or by changing the principal quantum number $n$ of the
excited Rydberg state. As the van der Waals interaction scales
with $n^{11}$ a reduction of the excitation rates was observed for
increasing $n$. A related interaction, the resonant dipole-dipole
interaction, has been investigated in the last years
\cite{Anderson:1998, Vogt:2006}, usually tuned with an electric
field but also with a magnetic field \cite{Afrousheh:2006}.
Many-body effects between some few atoms due to this interaction
have been spectroscopically resolved \cite{Mourachko:1998} and
their dependence on dimensionality was studied
\cite{Carroll:2006}. Recently, the first coherent excitations of
non-interacting ultracold atoms into a Rydberg state have been
achieved with the use of STIRAP sequences
\cite{Cubel:2005,Deiglmayr:2006}. Incoherent interactions could
result in a decay to different Rydberg states, ionization or state
changes by black body radiation.

In this paper we report on coherent Rydberg excitation of
magnetically trapped ultracold atoms in the strong blockade
regime. In this regime the excitation is strongly suppressed
compared to the non-interacting case and limited to a maximum
value which is in our experiment one out of few thousand ground
state atoms. We confirm the collective nature of the coherent
excitation by the dependence of the collective Rabi frequency on
the square root of the mesoscopic system size. This size
dependence is generic for all mesoscopic quantum systems for which
the excitation is restricted to a single quantum, including
so-called `superatoms' recently discussed in the context of single
photon storage \cite{Vuletic:2006}.

%\section{introduction to mesoscopic quantum dynamics}

A single atom exposed to resonant excitation light coherently
oscillates with the single-atom Rabi frequency $\Omega_0$ between
the ground and excited state. An ensemble of $N$ non-interacting
atoms gives just $N$ times the single-atom Rabi-oscillation at
frequency $\Omega_0$. But if for all members of the ensemble the
interaction between atoms in the excited state is much stronger
than the linewidth of the excitation, the ensemble can carry only
one excitation. As the excitation can be located at any of the $N$
atoms this collective state is of the form:
\begin{equation}\label{eqn_collstate}
    \ket{\psi_e}=\frac{1}{\sqrt{N}}\sum_{i=1}^{N}\ket{g_1,g_2,g_3,...,e_i,...,g_N},
\end{equation}
where $g_k$ indicates an atom numbered $k$ in the ground state and
$e_i$ one atom $i$ in the excited state. Therefore, the ensemble
is excited collectively and oscillates with the collective Rabi
frequency $\sqrt{N}\Omega_0$ between the ground state and the
collectively excited state $\ket{\psi_e}$ \cite{Lukin:2001}. In
this sense the ensemble of $N$ atoms acts like a `superatom'
\cite{Vuletic:2006} with a transition dipole moment which is
enhanced by a factor $\sqrt{N}$ as compared to the individual
atoms.

\begin{figure}
 \includegraphics[width=70mm]{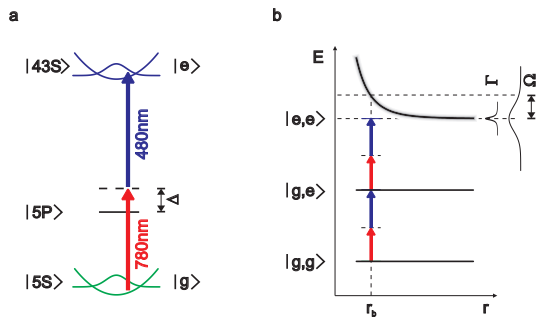}
 \caption{\label{vdw-WW}
 \textbf{a} Two-photon excitation of magnetically trapped atoms from the
  $5S_{1/2}, F=2, m_F=2$ state into the $43S_{1/2}, m_S=1/2$. Due to the large detuning $\Delta$,
  the three levels are reduced to an effective two-level system
  ($\ket{g}$,$\ket{e}$). \textbf{b} Molecular potential curves as
  a function of the relative coordinate $r$ (not to scale).
  The strong van der Waals interaction between the excited
  Rydberg states leads to a blockade effect for $\ket{e,e}$.
  Double excitation for distances smaller than the blockade radius
  $r_{\text{b}}$ is strongly suppressed. For two atoms the
  blockade radius is either determined by the linewidth $\Gamma$
  or
  the power broadening $\Omega$ (see Eq.\,\ref{eqn_blockade}).}

 \end{figure}

\begin{figure}
 %\psfrag{q}[c]{\textmu}
 \includegraphics[width=86mm]{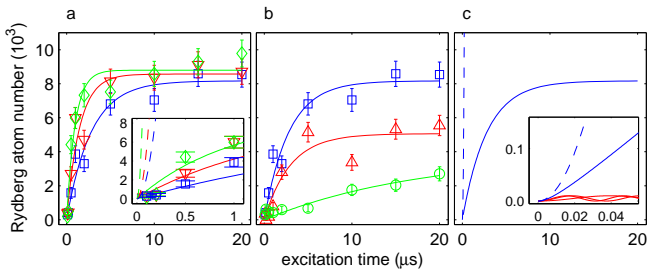}
 \caption{\label{Fig_Saturation}\textbf{a} The Rydberg atom number plotted versus excitation
 time for a high laser intensity ($\Omega_0=$ 210 kHz) and three values of the density of ground state atoms
 $n_{\text{g,0}}= $ (\unit[3.2$\times 10^{13}$] (\textcolor{green}{$\diamond$}),
 \unit[6.6$\times 10^{12}$] (\textcolor{red}{$\triangledown$}),
 \unit[2.8$\times 10^{12}$] (\textcolor{blue}{$\square$})) {cm$^{-3}$}. \textbf{b} The Rydberg atom number plotted versus excitation
 time for a low atom density ($n_{\text{g,0}}=2.8 \times 10^{12}\text{cm}^{-3}$) and three values of laser intensity
 $\Omega_0= $ (\unit[210] (\textcolor{blue}{$\square$}), \unit[93] (\textcolor{red}{$\vartriangle$}),
 \unit[42] (\textcolor{green}{$\circ$}) ) kHz.
  The solid curves are fits to the data with a
  simple exponential saturation curve. The inset in \textbf{a} shows a magnification of the
  data in contrast with the calculated Rabi oscillation (dashed) assuming negligible
  interactions.
  Fig. \textbf{c} shows a schematic of the excitation dynamics in an inhomogeneous sample. Many oscillating
  `superatoms'
   (shown with exaggerated amplitudes in \textcolor{red}{red}) add up to an integrated staturation curve (\textcolor{blue}{blue}). This curve falls
    behind the noninteracting case (shown for short times dashed in \textcolor{blue}{blue}) on a time scale of less than \unit[50]{ns}.
}
 \end{figure}

The so-called blockade radius is defined as the interatomic
distance where the interaction energy becomes equal to the
linewidth of the excitation, which is in our experiments dominated
by power-broadening (Fig.\,\ref{vdw-WW}b). We define the strong
blockade regime by a blockade radius significantly larger than the
mean interatomic distance, i.e. $N\gg 1$. In our experiments the
sample size is larger than the blockade radius, we therefore model
the sample by an ensemble of `superatoms'. Additionally in our
system, the density of ground state atoms and by this $N$, the
atom number per `superatom', is inhomogeneous. With it the
collective Rabi frequency is inhomogeneously distributed and the
local oscillations add up to a total population that for short
times increases quadratically. But after a very short time the
Rydberg population shows a linear increase that falls behind the
quadratic increase without interaction (see
Fig.\,\ref{Fig_Saturation}c). This time is related to the inverse
maximum collective Rabi frequency in the sample. For all
experimental conditions shown in this paper this time is shorter
than \unit[50]{ns}. For longer excitation times, the excitation in
the strong blockade regime can be distinguished from single-atom
behaviour by a strong suppression of excitation (see inset
Fig.\,\ref{Fig_Saturation}a). The time scale for the subsequent
linear increase is proportional to the inverse of the averaged
collective Rabi frequency $\sqrt{N_{\text{mean}}}\Omega_0$. Due to
the inhomogeneity the excited state population reaches after a
time, that is also related to
$(\sqrt{N_{\text{mean}}}\Omega_0)^{-1}$, a constant saturation
value that is determined by the number of `superatoms' in the
sample. This number shows, as it will be explained below, a
scaling with $\Omega_0$ and the density of ground state atoms,
which is characteristic for the underlying blockade mechanism.

%\section{experimental methods}

In our experiments we use magnetically trapped Rubidium atoms,
evaporatively cooled to a few \textmu K. We investigated the
excitation dynamics and observe full saturation for a large range
of densities and excitation rates. The observed scaling of the
initial increase and the saturation population with density and
Rabi frequency provides evidence of coherent collective excitation
as predicted by the `superatom' model.

We start with a sample of $N_{\text{g}}=1.5\times 10^7$
$^{87}\text{Rb}$ atoms in the $5S_{1/2}, F=2, m_F=2$ state at a
temperature of \unit[3.4]{\textmu K} and a Gaussian density
distribution with a peak value $n_{\text{g,0}}$ of
\unit[8.2$\times 10^{13}$]{cm$^{-3}$} in a specialized setup for
Rydberg experiments \cite{Loew:2007}. The excitation to the
$43S_{1/2}$ Rydberg state is done in a Ioffe-Pritchard-type trap
with a two-photon transition via the $5P_{3/2}$ state with a
detuning $\Delta$ to the blue by \unit[478]{MHz} (see
Fig.\,\ref{vdw-WW}a). We choose an $S$ state as it has only one
repulsive branch in its molecular potential (see
Fig.\,\ref{vdw-WW}b), whereas higher $l$ states typically have
repulsive and attractive branches and are subject to enhanced ion
formation. Resonant dipole-dipole interaction due to the
dominating transition $43S+43S$ $\rightarrow$ $42P+43P$ is
negligible for this experiment \cite{Li:2005}. For the $5S-5P$
transition the Rabi frequency $\Omega_1$ was determined by
Autler-Townes splitting at higher intensities
\cite{Grabowski:2006}. In the current experiment, $\Omega_1$ is
varied from \unit[2.0]{MHz} to \unit[9.7]{MHz}. For the upper
transition we estimate a Rabi frequency $\Omega_2$ of
\unit[21]{MHz} from our calculation of the dipole matrix element.
This gives a two-photon Rabi frequency
$\Omega_0=\Omega_1\Omega_2/(2\Delta)$ of up to \unit[210]{kHz}.
Due to the large detuning from the lower transition, the change in
the density and momentum distribution due to absorption of photons
is negligible \cite{Loew:2007}. The alignment of the excitation
laser beams to the offset field of the magnetic trap together with
the adjustment of polarizations makes it possible to preserve the
magnetic moment and avoid energy shifts due to magnetic fields
\cite{Loew:2007}. The waists of the Gaussian laser beams are large
compared to the $1/e^2$-radius of the sample and the Rabi
frequency is almost constant over the sample \cite{Loew:2007}.

During the experiment the excitation lasers are switched on for an
excitation time $\tau$, which is varied between \unit[100]{ns} and
\unit[20]{\textmu s}. The longest excitation time is shorter than
the \unit[100]{\textmu s} lifetime of the 43$S$ state
\cite{Gallagher:1994,Loew:2007}. Although the thermal motion of
the ground state atoms is frozen out on the time scale of the
excitation, attractive interaction between the Rydberg atoms can
lead to collisions and ionization within this time scale
\cite{Li:2005, Knuffman:2006}. To avoid all effects of ions and
electrons on the Rydberg atoms we chose a Rydberg state with
repulsive van der Waals interaction ($C_6=-1.7\times 10^{19}$ a.u.
\cite{Singer:2005}) and applied an electric field of
\unitfrac[200]{V}{m} strength during the excitation. With this
field enhanced ionization by means of trapped electrons is
suppressed \cite{Li:2004}, which would otherwise limit the
lifetime \cite{Robinson:2000}. Possibly produced ions are
extracted from the sample within a time of \unit[400]{ns}, which
is shorter than the time scales of the interactions of interest
here. We resolved a density dependent blue shift of the excitation
spectral line, which was not observed without field. This blue
shift is a clear evidence for repulsive interaction whereas the
electric field of charged particles would shift the spectroscopic
lines to the red since the Stark shift is negative for this state.
In the described experiments the excitation lasers are tuned to
resonance, which was determined from an excitation spectrum. This
was done with very low laser power and thus low Rydberg densities
at which no line shift and broadening caused by interactions are
observed. The linewidth of excitation was measured to be smaller
than \unit[130]{kHz} on the microsecond time scale. This was done
in an echo-type experiment where the excitation dynamics could be
reversed \cite{Raitzsch:2007}.

Directly after the excitation pulse the excited Rydberg atoms are
field-ionized and the ions detected with a microchannel plate
(MCP) detector. The MCP was calibrated and the linearity over the
used range verified.

Additionally to the variation of excitation times, we changed the
two-photon Rabi frequency $\Omega_0$ by changing the power of the
\unit[780]{nm} excitation laser as well as the initial peak
density $n_{\text{g,0}}$ of the ground state atoms. The latter is
done by adiabatically transferring up to \unit[97]{\%} of the
atoms with a 6.8 GHz microwave Landau-Zener sweep of variable
duration to the untrapped $5S_{1/2},F=1, m_F=1$ state. Due to the
large detuning, this state is not affected by the excitation
light. With this technique we can vary the peak density with
almost no change in temperature and shape of the density
distribution. Every excitation and detection of the field-ionized
Rydberg atoms is followed by a \unit[20]{ms} time-of-flight of the
remaining atoms. We take an absorption image of the remaining
ground state atoms from which we obtain the ground state atom
number. With the temperature and the trapping potential we
calculate the density distribution.

%\section{initial excitation dynamics\label{sec_initial}}

Figure\,\ref{Fig_Saturation} shows the typical excitation dynamics
for  three different densities of the ground state atoms (a) and
three different Rabi frequencies (b). Two features are prominent
in the figure: initially a linear increase with time and a
saturation to a constant value. In contrast to previous
experiments at considerably lower densities, the dynamics has to
be described by full-quantum calculations \cite{Robicheaux:2005}
rather than a mean-field model \cite{Tong:2004}. However for the
following investigations the excitation dynamics curves were
fitted with a simple exponential saturation function of the form
\begin{equation}
N_{\text{R}}(\tau)=N_{\text{sat}}(1 - e^{-R \tau/N_{\text{sat}}
}),
\end{equation}
with the Rydberg atom number $N_{\text{R}}(\tau)$ after the
excitation time $\tau$, since we are mainly interested in the
scaling of the saturation Rydberg atom number $N_{\text{sat}}$ and
the initial slope $R$. The inset in Fig.\,\ref{Fig_Saturation}a
contrasts the Rabi oscillation of non-interacting atoms with our
measurement and demonstrates the strong blockade of excitation
already in the initial linear increase.

 \begin{figure}
 \includegraphics[width=86mm]{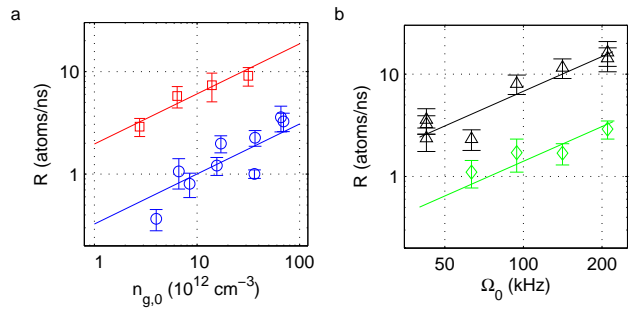}
 \caption{\label{Fig_slope}\textbf{a} Dependence of the initial increase $R$ of the
 Rydberg atom number on the density of ground state atoms $n_{\text{g,0}}$ for high (\textcolor{red}{$\square$})
 and low (\textcolor{blue}{$\circ$}) Rabi frequency $\Omega_0= $(210, 42)kHz. \textbf{b} Dependence of $R$ on the
 excitation rate $\Omega_0$ for high (\textcolor{black}{$\vartriangle$}) and low (\textcolor{green}{$\diamond$}) atom
 density $n_{\text{g,0}}= $(7.2$\times10^{13}$, 2.8$\times10^{12}$) cm$^{-3}$.
 %The red dataset (\textcolor{red}{$\square$}) contains the measurements shown in Fig. \ref{Fig_Saturation}a,
%   the green dataset (\textcolor{green}{$\diamond$}) contains the measurements shown in Fig. \ref{Fig_Saturation}b.
   The lines are the result of a power-law fit to the whole dataset in \textbf{a} and \textbf{b} of the form
   $R\propto n_{\text{g,0}}^{a} \Omega_0^{b}$ which gives an exponent for the $n_{\text{g,0}}$-dependence
    of $a=0.49\pm$0.06 which is in excellent agreement with
    the expected $\sqrt{n_{\text{g,0}}}$-scaling for collective excitation.
    The fitted exponent for the $\Omega_0$-scaling is
    $b=1.1\pm$0.1, which
    is in good agreement with a linear scaling with $\Omega_0$ for coherent
    excitation.}
 \end{figure}

Figure\,\ref{Fig_slope}a shows the scaling of the initial slope
$R$ of the excitation with the density of ground state atoms. In
the simplest model, the density $n_{\text{g,0}}$ is proportional
to $N$ (see Eq. \ref{eqn_blockade}). For non-interacting Rydberg
atoms, the excited fraction would be independent of the atom
number and $R$ would scale linearly with $N$. In contrast, $R$
shows a $\sqrt{N}$-scaling, a clear evidence for a collective
excitation. Furthermore, we investigated the scaling with the Rabi
frequency $\Omega_0$ by altering the intensity of the
\unit[780]{nm} laser. The initial excitation rate of
non-interacting atoms or strongly damped (i.e. incoherent)
excitation would scale with $\Omega_0^2$. The linear scaling
($R\propto \Omega_0$) being determined from Fig.\,\ref{Fig_slope}b
is an evidence for coherent excitation of the Rydberg atoms. The
combined $\sqrt{N}\Omega_0$-dependence is a clear evidence for
local coherent collective Rabi oscillations within a cloud with
spatially inhomogeneous density.

 \begin{figure}
 \includegraphics[width=86mm]{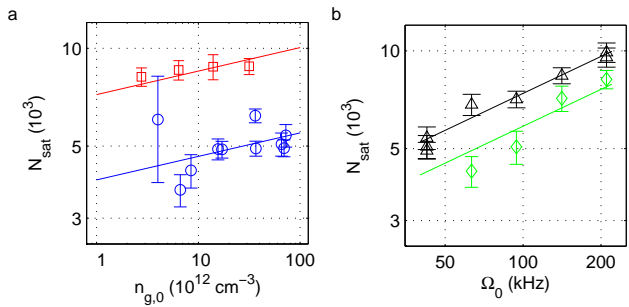}
 \caption{\label{Fig_satvalue}\textbf{a} Dependence of the saturation number of Rydberg
  atoms $N_{\text{sat}}$ on the density of ground state atoms  $n_{\text{g,0}}$ for high (\textcolor{red}{$\square$}) and
  low (\textcolor{blue}{$\circ$}) Rabi frequency $\Omega_0=$(210, 42) kHz. \textbf{b} Dependence of $N_{\text{sat}}$ on the
  Rabi frequency $\Omega_0$ for high (\textcolor{black}{$\vartriangle$}) and low (\textcolor{green}{$\diamond$}) atom
  density $n_{\text{g,0}} = $ (7.2$\times10^{13}$, 2.8$\times10^{12}$) cm$^{-3}$ .
 %The red dataset (\textcolor{red}{$\square$}) contains the measurements shown in Fig. \ref{Fig_Saturation}a,
%   the green dataset (\textcolor{green}{$\diamond$}) contains the measurements shown in Fig. \ref{Fig_Saturation}b.
   The lines are the result of a power-law fit to the whole dataset in \textbf{a} and \textbf{b} of the form
   $N_{\text{sat}}\propto n_{\text{g,0}}^{c} \Omega_0^{d}$ which gives an exponent for the
   $n_{\text{g,0}}$-dependence of $c=0.07\pm$0.02, which is in agreement with
    the expected independence from $n_{\text{g,0}}$ for strong blockade.
    The fitted exponent for the $\Omega_0$-scaling is
    $d=0.38\pm$0.04, which is in excellent agreement with the expected $\Omega_0^{2/5}$-scaling for a collective van der Waals blockade.}
 \end{figure}

%\section{saturation regime}

The blockade radius depends on the van der Waals interaction
strength and the linewidth of the excitation. In our excitation
scheme, using cw lasers and large detuning from the intermediate
state, the linewidth is dominated by the Rabi frequency
$\Omega_0$. In the simplest model, we estimate the blockade radius
as the distance at which the van der Waals interaction $C_6/r^6$
equals the power-broadened linewidth $\hbar \Omega_0$ except for a
geometric factor that includes the arrangement of `superatoms'
(see Fig.\,\ref{vdw-WW}b):
\begin{equation}\label{eqn_blockade}
    r_{\text{b}}\propto[C_6/(\hbar\Omega_0)]^{1/6}.
\end{equation}
As the saturation density of Rydberg atoms is proportional to
$r_{\text{b}}^{-3}$ and the saturation number of Rydberg atoms
$N_{\text{sat}}$ is proportional to their density,
$N_{\text{sat}}$ is expected proportional to $\sqrt{\Omega_0}$ and
independent of the density of ground state atoms. In reasonable
agreement with this expectation we observe a very weak
$n_{\text{g,0}}$-dependence in the saturation number of Rydberg
atoms as shown in Fig.\,\ref{Fig_satvalue}a although we change the
ground state density by more than an order of magnitude. The
average atom number per `superatom'
$N_{\text{mean}}=N_{\text{g}}/N_{\text{sat}}$ is between 65 and
2500, while we expect $N$ to be one order of magnitude higher in
the centre of the cloud. Therefore in our experimental setup,
using this Rydberg state, a direct observation of single-atom Rabi
oscillations is not possible since we would have to reduce the
density by a factor of 2500 to about \unit[$10^{10}$]{cm$^{-3}$}.
This corresponds to a reduction to 4000 atoms which is not
possible in a controlled way.

In Fig.\,\ref{Fig_satvalue}b the saturation value of the Rydberg
atom number is plotted against the single-atom Rabi frequency. The
observed dependence is close to the expected
$\sqrt{\Omega_0}$-scaling which is characteristic for van der
Waals interaction. Note that in Eq.\,\ref{eqn_blockade} additional
$N$-dependent terms are expected e.g. a
$\sqrt{N}\Omega_0$-behaviour of the collective Rabi frequency.
This closer consideration gives a scaling of $N_{\text{sat}}$ with
$n_{\text{g,0}}^{1/5}\Omega_0^{2/5}$ which is in even better
agreement with the experimental observation. Other density
dependent effects like number of next neighbours are currently
under further theoretical investigation.

%\section{Conclusions}

To conclude, we have found evidence for mesoscopic quantum
dynamics of frozen Rydberg gases in the strong blockade regime.
Mesoscopic size effects on the coherent evolution have been
identified for up to a few thousand atoms per mesoscopic unit.
This became possible by narrowband excitation of magnetically
trapped atoms at temperatures of a few microKelvin and variable
densities. Analogous size effects are expected in other mesoscopic
systems carrying a single excitation quantum only, like an exciton
in a quantum dot or a dark state polariton excited by a single
photon in an ensemble of atoms \cite{Fleischhauer:2005}. In the
latter the time scale for the coherent evolution of the mesoscopic
ensemble also speeds up by a $\sqrt{N}$ factor, which is an
important factor for quantum repeaters enabling long-distance
quantum communication \cite{Duan:2001}. The demonstrated
scalability of the system will enable studies of size dependent
quantum correlations and decoherence effects in strongly
interacting non-equilibrium situations. In future experiments
using Bose-Einstein condensates, phase sensitive measurements
beyond mean field might become possible.

%\begin{acknowledgments}
We would like to thank Helmar Bender who set up the
\unit[6.8]{GHz} source and Thierry Lahaye for proof-reading. We
acknowledge financial support from the Deutsche
Forschungsgemeinschaft within the SFB/TRR21, SFB407, SPP116 and
under the contract PF 381/4-1, U.R. acknowledges support from the
Landesgraduiertenf\"{o}rderung Baden-W\"{u}rttemberg.\\
%\end{acknowledgments}


\begin{thebibliography}{10}
\providecommand{\bibnamefont}[1]{#1}
\providecommand{\bibfnamefont}[1]{#1}
\providecommand{\bibinfo}[2]{#2}

\bibitem{Raimond:1981}
\bibinfo{author}{\bibfnamefont{J.}~\bibnamefont{Raimond}},
  \bibinfo{author}{\bibfnamefont{G.}~\bibnamefont{Vitrant}}, \bibnamefont{and}
  \bibinfo{author}{\bibfnamefont{S.}~\bibnamefont{Haroche}},
  \bibinfo{journal}{J. Phys. B} \textbf{\bibinfo{volume}{14}},
  \bibinfo{pages}{L655} (\bibinfo{year}{1981}).

\bibitem{Tong:2004}
\bibinfo{author}{\bibfnamefont{D.}~\bibnamefont{{Tong}}}, \emph{et~al.},
  \bibinfo{journal}{Phys. Rev. Lett.}
  \textbf{\bibinfo{volume}{93}}(\bibinfo{number}{6}), \bibinfo{pages}{063001}
  (\bibinfo{year}{2004}).

\bibitem{Singer:2004}
\bibinfo{author}{\bibfnamefont{K.}~\bibnamefont{{Singer}}}, \emph{et~al.},
  \bibinfo{journal}{Phys. Rev. Lett.}
  \textbf{\bibinfo{volume}{93}}(\bibinfo{number}{16}), \bibinfo{pages}{163001}
  (\bibinfo{year}{2004}).

\bibitem{Liebisch:2005}
\bibinfo{author}{\bibfnamefont{T.~C.} \bibnamefont{{Liebisch}}},
  \bibinfo{author}{\bibfnamefont{A.}~\bibnamefont{{Reinhard}}},
  \bibinfo{author}{\bibfnamefont{P.~R.} \bibnamefont{{Berman}}},
  \bibnamefont{and}
  \bibinfo{author}{\bibfnamefont{G.}~\bibnamefont{{Raithel}}},
  \bibinfo{journal}{Phys. Rev. Lett.}
  \textbf{\bibinfo{volume}{95}}(\bibinfo{number}{25}), \bibinfo{pages}{253002}
  (\bibinfo{year}{2005}).

\bibitem{Anderson:1998}
\bibinfo{author}{\bibfnamefont{W.~R.} \bibnamefont{{Anderson}}},
  \bibinfo{author}{\bibfnamefont{J.~R.} \bibnamefont{{Veale}}},
  \bibnamefont{and} \bibinfo{author}{\bibfnamefont{T.~F.}
  \bibnamefont{{Gallagher}}}, \bibinfo{journal}{Phys. Rev. Lett.}
  \textbf{\bibinfo{volume}{80}}, \bibinfo{pages}{249} (\bibinfo{year}{1998}).

\bibitem{Vogt:2006}
\bibinfo{author}{\bibfnamefont{T.}~\bibnamefont{{Vogt}}}, \emph{et~al.},
  \bibinfo{journal}{Phys. Rev. Lett.}
  \textbf{\bibinfo{volume}{97}}(\bibinfo{number}{8}), \bibinfo{pages}{083003}
  (\bibinfo{year}{2006}).

\bibitem{Afrousheh:2006}
\bibinfo{author}{\bibfnamefont{K.}~\bibnamefont{Afrousheh}}, \emph{et~al.},
  \bibinfo{journal}{Phys. Rev. A}
  \textbf{\bibinfo{volume}{73}}(\bibinfo{number}{6}), \bibinfo{eid}{063403}
  (\bibinfo{year}{2006}).

\bibitem{Mourachko:1998}
\bibinfo{author}{\bibfnamefont{I.}~\bibnamefont{{Mourachko}}}, \emph{et~al.},
  \bibinfo{journal}{Phys. Rev. Lett.} \textbf{\bibinfo{volume}{80}},
  \bibinfo{pages}{253} (\bibinfo{year}{1998}).

\bibitem{Carroll:2006}
\bibinfo{author}{\bibfnamefont{T.~J.} \bibnamefont{{Carroll}}},
  \bibinfo{author}{\bibfnamefont{S.}~\bibnamefont{{Sunder}}}, \bibnamefont{and}
  \bibinfo{author}{\bibfnamefont{M.~W.} \bibnamefont{{Noel}}},
  \bibinfo{journal}{PRA} \textbf{\bibinfo{volume}{73}}(\bibinfo{number}{3}),
  \bibinfo{pages}{032725} (\bibinfo{year}{2006}).

\bibitem{Cubel:2005}
\bibinfo{author}{\bibfnamefont{T.}~\bibnamefont{Cubel}}, \emph{et~al.},
  \bibinfo{journal}{PRA} \textbf{\bibinfo{volume}{72}}, \bibinfo{pages}{023405}
  (\bibinfo{year}{2005}).

\bibitem{Deiglmayr:2006}
\bibinfo{author}{\bibfnamefont{J.}~\bibnamefont{{Deiglmayr}}}, \emph{et~al.},
  \bibinfo{journal}{Opt. Commun.} \textbf{\bibinfo{volume}{264}},
  \bibinfo{pages}{293} (\bibinfo{year}{2006}).

\bibitem{Jaksch:2000}
\bibinfo{author}{\bibfnamefont{D.}~\bibnamefont{{Jaksch}}}, \emph{et~al.},
  \bibinfo{journal}{Phys. Rev. Lett.} \textbf{\bibinfo{volume}{85}},
  \bibinfo{pages}{2208} (\bibinfo{year}{2000}).

\bibitem{Lukin:2001}
\bibinfo{author}{\bibfnamefont{M.~D.} \bibnamefont{{Lukin}}}, \emph{et~al.},
  \bibinfo{journal}{Phys. Rev. Lett.}
  \textbf{\bibinfo{volume}{87}}(\bibinfo{number}{3}), \bibinfo{pages}{037901}
  (\bibinfo{year}{2001}).

\bibitem{Vuletic:2006}
\bibinfo{author}{\bibfnamefont{V.}~\bibnamefont{Vuletic}},
  \bibinfo{journal}{Nature Physics} \textbf{\bibinfo{volume}{2}},
  \bibinfo{pages}{801} (\bibinfo{year}{2006}).

\bibitem{Loew:2007}
\bibinfo{author}{\bibfnamefont{R.}~\bibnamefont{L{\"o}w}}, \emph{et~al.},
  \bibinfo{journal}{arXiv:0706.2639}  (\bibinfo{year}{2007}).

\bibitem{Li:2005}
\bibinfo{author}{\bibfnamefont{W.}~\bibnamefont{{Li}}},
  \bibinfo{author}{\bibfnamefont{P.~J.} \bibnamefont{{Tanner}}},
  \bibnamefont{and} \bibinfo{author}{\bibfnamefont{T.~F.}
  \bibnamefont{{Gallagher}}}, \bibinfo{journal}{Phys. Rev. Lett.}
  \textbf{\bibinfo{volume}{94}}(\bibinfo{number}{17}), \bibinfo{pages}{173001}
  (\bibinfo{year}{2005}).

\bibitem{Grabowski:2006}
\bibinfo{author}{\bibfnamefont{A.}~\bibnamefont{Grabowski}}, \emph{et~al.},
  \bibinfo{journal}{Fortschr. Phys.} \textbf{\bibinfo{volume}{54}},
  \bibinfo{pages}{765} (\bibinfo{year}{2006}).

\bibitem{Gallagher:1994}
\bibinfo{author}{\bibfnamefont{T.~F.} \bibnamefont{Gallagher}},
  \emph{\bibinfo{title}{Rydberg Atoms}} (\bibinfo{publisher}{Cambrige
  University Press, Cambridge}, \bibinfo{year}{1994}).

\bibitem{Knuffman:2006}
\bibinfo{author}{\bibfnamefont{B.}~\bibnamefont{{Knuffman}}} \bibnamefont{and}
  \bibinfo{author}{\bibfnamefont{G.}~\bibnamefont{{Raithel}}},
  \bibinfo{journal}{PRA} \textbf{\bibinfo{volume}{73}}(\bibinfo{number}{2}),
  \bibinfo{pages}{020704(R)} (\bibinfo{year}{2006}).

\bibitem{Singer:2005}
\bibinfo{author}{\bibfnamefont{K.}~\bibnamefont{{Singer}}},
  \bibinfo{author}{\bibfnamefont{J.}~\bibnamefont{{Stanojevic}}},
  \bibinfo{author}{\bibfnamefont{M.}~\bibnamefont{{Weidem{\"u}ller}}},
  \bibnamefont{and}
  \bibinfo{author}{\bibfnamefont{R.}~\bibnamefont{{C{\^o}t{\'e}}}},
  \bibinfo{journal}{J. Phys. B: At., Mol. Opt. Phys.}
  \textbf{\bibinfo{volume}{38}}, \bibinfo{pages}{295} (\bibinfo{year}{2005}).

\bibitem{Li:2004}
\bibinfo{author}{\bibfnamefont{W.}~\bibnamefont{{Li}}}, \emph{et~al.},
  \bibinfo{journal}{PRA} \textbf{\bibinfo{volume}{70}}(\bibinfo{number}{4}),
  \bibinfo{pages}{042713} (\bibinfo{year}{2004}).

\bibitem{Robinson:2000}
\bibinfo{author}{\bibfnamefont{M.~P.} \bibnamefont{{Robinson}}}, \emph{et~al.},
  \bibinfo{journal}{Phys. Rev. Lett.} \textbf{\bibinfo{volume}{85}},
  \bibinfo{pages}{4466} (\bibinfo{year}{2000}).

\bibitem{Raitzsch:2007}
\bibinfo{author}{\bibfnamefont{U.}~\bibnamefont{Raitzsch}}, \emph{et~al.},
  \bibinfo{journal}{in preparation}  (\bibinfo{year}{2007}).

\bibitem{Robicheaux:2005}
\bibinfo{author}{\bibfnamefont{F.}~\bibnamefont{{Robicheaux}}}
  \bibnamefont{and} \bibinfo{author}{\bibfnamefont{J.~V.}
  \bibnamefont{{Hern{\'a}ndez}}}, \bibinfo{journal}{PRA}
  \textbf{\bibinfo{volume}{72}}(\bibinfo{number}{6}), \bibinfo{pages}{063403}
  (\bibinfo{year}{2005}).

\bibitem{Fleischhauer:2005}
\bibinfo{author}{\bibfnamefont{M.}~\bibnamefont{Fleischhauer}},
  \bibinfo{author}{\bibfnamefont{A.}~\bibnamefont{Imamoglu}}, \bibnamefont{and}
  \bibinfo{author}{\bibfnamefont{J.~P.} \bibnamefont{Marangos}},
  \bibinfo{journal}{Rev. Mod. Phys.}
  \textbf{\bibinfo{volume}{77}}(\bibinfo{number}{2}), \bibinfo{eid}{633}
  (\bibinfo{year}{2005}).

\bibitem{Duan:2001}
\bibinfo{author}{\bibfnamefont{L.-M.} \bibnamefont{{Duan}}},
  \bibinfo{author}{\bibfnamefont{M.~D.} \bibnamefont{{Lukin}}},
  \bibinfo{author}{\bibfnamefont{J.~I.} \bibnamefont{{Cirac}}},
  \bibnamefont{and} \bibinfo{author}{\bibfnamefont{P.}~\bibnamefont{{Zoller}}},
  \bibinfo{journal}{\nat} \textbf{\bibinfo{volume}{414}}, \bibinfo{pages}{413}
  (\bibinfo{year}{2001}).

\end{thebibliography}
\end{document}